\documentclass{elsart}

\usepackage{graphicx}

\usepackage{amssymb}

\begin{document}

\begin{frontmatter}



\title{$H$-$T$ phase diagram and the nature of Vortex-glass phase in a
quasi two-dimensional superconductor: Sn metal layer sandwiched between
graphene sheets}

\author[SUNY]{Masatsugu Suzuki}
\ead{suzuki@binghamton.edu}
\author[SUNY]{Itsuko S. Suzuki}
\author[OSAKA]{J\"{u}rgen Walter}
\address[SUNY]{Department of Physics, State University of New York at Binghamton, 
Binghamton, New York 13902-6016, U.S.A.}
\address[OSAKA]{Department of Materials Science and Processing, Graduate School 
of Engineering, Osaka University, 2-1, Yamada-oka, Suita, 565-0879, 
JAPAN}

\begin{abstract}
The magnetic properties of a quasi two-dimensional superconductor, Sn-metal
graphite (MG), are studied using DC and AC magnetic susceptibility.  Sn-MG
has a unique layered structure where Sn metal layer is sandwiched between
adjacent graphene sheets.  This compound undergoes a superconducting
transition at $T_{c}$ = 3.75 K at $H$ = 0.  The $H$-$T$ diagram of Sn-MG is
similar to that of a quasi two-dimensional superconductors.  The phase
boundaries of vortex liquid, vortex glass, and vortex lattice phase merge
into a multicritical point located at $T^{*}$ = 3.4 K and $H^{*}$ = 40 Oe. 
There are two irreversibility lines denoted by $H_{gl}$ (de
Almeida-Thouless type) and $H_{gl^{\prime}}$ (Gabay-Toulouse type),
intersecting at $T_{0}^{\prime}$= 2.5 K and $H_{0}^{\prime}$ = 160 Oe.  The
nature of slow dynamic and nonlinearity of the vortex glass phase is
studied.
\end{abstract}

\end{frontmatter}

\section{\label{intro}Introduction}
The existence of vortex glass phase in the mixed state of quasi
two-dimensional (2D) high-$T_{c}$ superconductors such as
YBa$_{2}$Cu$_{3}$O$_{7}$ in the presence of magnetic field ($H$) attracts
considerable attention \cite{Blatter1994,Gammel1998,Menon2002}.  M\"{u}ller
et al.  \cite{Muller1987} have pointed out for the first time the existence
of the glassy phase in La$_{2}$CuO$_{4-y}$:Ba.  They have shown that the
boundary between the vortex liquid phase and the vortex glass phase
corresponds to the de Almeida-Thouless (AT) \cite{Almeida1978} line for the
spin glass behavior.  Since then there have been a number of experimental
works on the vortex glass phase in quasi 2D superconductors. 
Experimentally it has been confirmed that the $H$-$T$ diagram consists of
the vortex glass, vortex lattice (Abrikosov lattice), and vortex liquid
phases.  The boundary between the vortex lattice phase and the vortex
liquid phase (the line $H_{al}$) is of the first order, while the boundary
between the vortex lattice phase and the vortex glass phase (the line
$H_{ag}$) and the boundary between the vortex glass phase and the vortex
liquid phase (the line $H_{gl}$) are of the second order.  These boundaries
merge into a multicritical point at $T$ =$T^{*}$ and $H$ = $H^{*}$.

Sn-metal graphite (MG) constitutes a novel class of materials having unique
layered structures.  This system can be prepared by reduction of an
acceptor-type SnCl$_{2}$ graphite intercalation compound (GIC) as a
precursor material.  Ideally, the staging structure of Sn-MG would be the
same as that of SnCl$_{2}$ GIC. Sn layer is sandwiched between adjacent
graphene sheets.  These sandwiched structures are periodically stacked
along the $c$ axis perpendicular to the basal plane of graphene sheets. 
Sn-MG undergoes a superconducting transition at a critical temperature
$T_{c}$ (= 3.75 K), which is close to that of bulk Sn \cite{Maxwell1952}. 
The superconductivity occurs mainly in Sn metal layers, forming a quasi 2D
superconductor (type-II).

In this paper we report results on the magnetic properties of Sn-MG from
measurements of DC and AC magnetic susceptibility.  Three methods are used
to determine the $H$-$T$ diagram: (i) the $T$ dependence of the
magnetization ($M$) in various states such as the FC (field-cooled) state,
ZFC (zero-field cooled) state, IR (isothermal remnant ) state, and TR
(thermoremnant) state, (ii) the $T$, $H$, $h$, and $f$ dependence of the
dispersion ($\Theta_{1}^{\prime}/h$ or $\chi^{\prime}$) and absorption
($\Theta_{1}^{\prime\prime}/h$ or $\chi^{\prime\prime}$) of the AC magnetic
susceptibility, where $h$ and $f$ are the amplitude and frequency of the AC
magnetic field, respectively, and (iii) the $H$ dependence of $M$.  We find
that there are four lines ($H_{ag}$, $H_{gl}$, $H_{gl^{\prime}}$ and
$H_{al}$) in the $H$-$T$ phase diagram, separating vortex liquid, vortex
glass, and vortex lattice phase.  These lines merge at the multicritical
point located at $T^{*}$ = 3.4 K and $H^{*}$ = 40 Oe.  The lines $H_{gl}$
and $H_{gl^{\prime}}$ are irreversibility lines, intersecting at a critical
point at $T_{0}$' = 2.5 K and $H_{0}^{\prime}$ = 160 Oe.  The line $H_{gl}$
has an AT-power law form,\cite{Almeida1978} while the line
$H_{gl^{\prime}}$ has a Gabay-Thouless (GT)-power law form
\cite{Gabay1981}.  The nature of possible slow dynamics and nonlinearity in
the vortex glass phase is examined from the measurement of
$\Theta_{1}^{\prime}/h$ and $\Theta_{1}^{\prime\prime}/h$ as a function of
$\omega$, $h$, $T$, and $H$.

\section{\label{exp}Experimental procedure}
The method of sample preparation in Sn-MG is similar to that in other MG's
previously reported
\cite{Walter1999,Walter2000a,Walter2000b,Walter2000c}. SnCl$_{2}$
graphite intercalation compound (GIC) samples as a precursor material, were
prepared by heating a mixture of natural graphite flakes (grade: RFL 99.9
S) from Kropfm\"{u}hl, Germany and an excess amount of SnCl$_{2}$.  The
reaction was continued for three days at 400 $^\circ$C in an ampoule filled
with chlorine gas.  The synthesis of Sn-MG was made by the reduction of
SnCl$_{2}$ GIC. SnCl$_{2}$ GIC samples were kept for two days in a solution
of lithium diphenylide in tetrahydrofuran (THF) at room temperature.  Then
the samples were filtered, rinsed by THF, and dried in air.  The structure
of Sn-MG thus obtained was examined by (00$L$) x-ray diffraction, and
bright field images and selected-area electron diffraction (SAED) (Hitachi
H-800 transmission electron microscope) operated at 200 kV. No pristine
SnCl$_{2}$ and its hydrolisation products were left in the Sn layers.  The
details of sample preparation and characterization in Sn-MG will be
reported elsewhere.

The Sn-MG sample consists of many small flakes.  Each flake has a
well-defined $c$ axis.  The sample in the present work is regarded as a
powdered sample with the $c$ axis randomly distributed in all directions,
since these flakes are randomly piled inside the sample capsule.  The
measurements of DC and AC magnetic susceptibility were carried out using a
SQUID magnetometer (Quantum Design MPMS XL-5).  Before setting up a sample
at 298 K, a remnant magnetic field was reduced to less than 3 mOe using an
ultra-low field capability option.  For convenience, hereafter this remnant
field is noted as the state $H$ = 0.  (i)\textit{ZFC and FC magnetic
susceptibility}: The sample was cooled from 298 to 1.9 K at $H$ = 0.  After
$H$ was applied at 1.9 K, the zero-field cooled magnetization ($M_{ZFC}$)
was measured with increasing $T$ from 1.9 to 6 K. The sample was kept at 20
K for 20 minutes in the presence of $H$.  Then the field cooled
magnetization ($M_{FC}$) was measured with decreasing $T$ from 6 to 1.9 K.
(ii) $M$-$H$ loop: The sample was cooled from 298 K to $T$ at $H = 0$.  The
measurement was carried out with varying $H$ from 0 to 3 kOe at $T$, from
$H$ = 3 to -3 kOe, and from $H = -3$ to 3 kOe.  (iii) $M_{ZFC}$ vs $H$: The
sample was cooled from 298 K to $T$ at $H = 0$.  The ZFC magnetization
$M_{ZFC}$ at $T$ was measured with increasing $H$ from 0 to 250 Oe.  (iv)
\textit{AC magnetic susceptibility}.  The dispersion
($\Theta_{1}^{\prime}/h$) and the absorption
($\Theta_{1}^{\prime\prime}/h$) was determined from the in-phase and out-of
phase first harmonic ($\omega$) component in the AC magnetization
measurement \cite{Suzuki2002a}.  The dispersion ($\Theta_{1}^{\prime}/h$)
and the absorption ($\Theta_{1}^{\prime\prime}/h$) were described by
$\Theta_{1}^{\prime}/h = \chi^{\prime} + 3\chi_{3}^{\prime}h^{2}/4 +
5\chi_{5}^{\prime}h^{4}/8 + \dots$ and $\Theta_{1}^{\prime\prime}/h =
\chi^{\prime\prime} + 3\chi_{3}^{\prime\prime}h^{2}/4 +
5\chi_{5}^{\prime\prime}h^{4}/8 + \cdots$ , where $\chi_{2n+1}^{\prime}$
and $\chi_{2n+1}^{\prime\prime}$ ($n = 1, 2, \cdots$) are the real and
imaginary parts of high-order nonlinear magnetic susceptibility,
respectively.  In the limit of $h = 0$, $\Theta_{1}^{\prime}/h$ and
$\Theta_{1}^{\prime\prime}/h$ coincide with the linear susceptibility
$\chi^{\prime}$ and $\chi^{\prime\prime}$, respectively.  Before the
measurement, the sample was cooled from 298 to 1.9 K at $H = 0$.  (i) The
$T$ dependence of $\Theta_{1}^{\prime}/h$ and $\Theta_{1}^{\prime\prime}/h$
was measured with increasing $T$ from 1.9 to 6 K in the presence of $H$. 
After the measurement, the sample was cooled from 6 to 1.9 K in the
presence of the same $H$.  Then the measurement was repeated at different
$H$.  (ii) The $h$ dependence was carried out at fixed $T$ by changing $h$
($0.01 \leq h \leq 4$ Oe) at $f = 1$ Hz.  (iii) The $f$ dependence was
carried out at fixed $T$ in the presence of $H$ by changing $f$ ($0.1 \leq
f \leq 1000$ Hz).

\section{\label{result}Result} 
\subsection{\label{resultA}$\chi_{FC}$ and $\chi_{ZFC}$ }

\begin{figure}
     \begin{center}
     \includegraphics*[width=8cm]{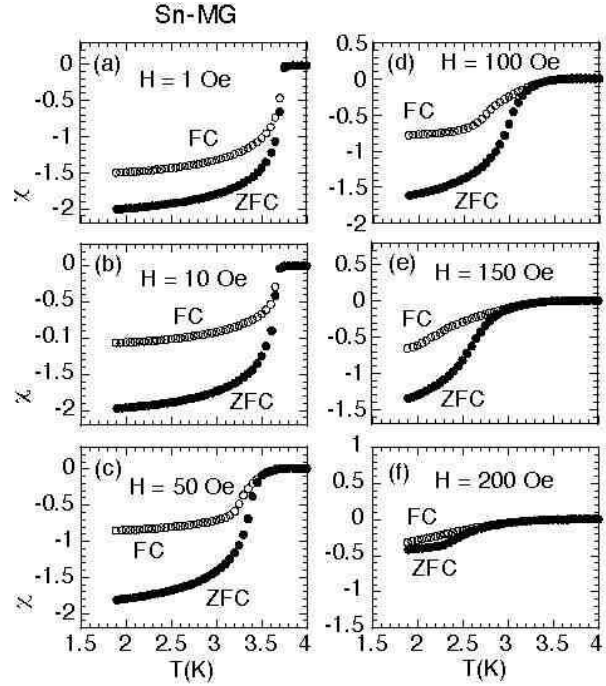}
     \end{center}
\caption{(a) - (f) $T$ dependence of $\chi_{FC}$ and $\chi_{ZFC}$ at low
$H$ ($1 \leq H \leq 200$ Oe) for Sn-MG.}
\label{fig:one}
\end{figure}

\begin{figure}
     \begin{center}
     \includegraphics*[width=8cm]{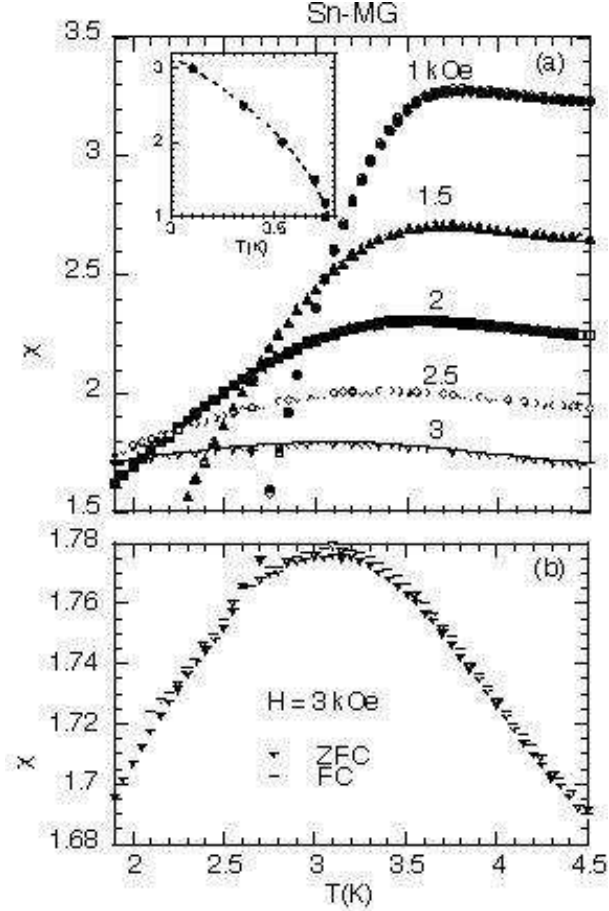}
     \end{center}
\caption{$T$ dependence of $\chi_{FC}$ and $\chi_{ZFC}$ at high $H$ ($1
\leq H \leq 3$ kOe) for Sn-MG. The inset shows the plot of the peak
temperature of $\chi_{FC}$ vs $T$ as a function of $H$.  The dotted line is
a least-squares fitting curve.  See the text for detail.  (b) $T$
dependence of $\chi_{FC}$ and $\chi_{ZFC}$ at $H$ = 3 kOe, which is a
blow-up of Fig.~\ref{fig:two}(a).}
\label{fig:two}
\end{figure}

Figure \ref{fig:one} shows the $T$ dependence of $\chi_{FC}$ and
$\chi_{ZFC}$ at low $H$.  The susceptibility $\chi_{ZFC}$ deviates from
$\chi_{FC}$ below a characteristic temperature for $1 \leq H \leq 200$ Oe,
suggesting the irreversibility line in the $H$-$T$ plane which separates
the reversible region from the irreversible region.  Figure 2 shows the $T$
dependence of $\chi_{FC}$ and $\chi_{ZFC}$ for $1 \leq H \leq 3$ kOe.  For
$H \geq 300$ Oe there is no noticeable difference between $\chi_{FC}$ and
$\chi_{ZFC}$ at any $T$.  The susceptibility ($\chi_{ZFC} = \chi_{FC}$)
exhibits a very broad peak, which shifts to the low-$T$ side as $H$
increases.  In the inset of Fig.~\ref{fig:two}(a) we plot these peak
temperatures as a function of $H$.  The value of $H$ (= $H_{p}$) for the
peak is much larger than that of a upper critical field $H_{c2}$ (whose
definition will be given in Sec.~\ref{disA}) at the same $T$, and it tends
to zero around $T_{c}$.  This result suggests that an antiferromagnetic
(AF) short-range order survives at high $H$ where the superconductivity
disappears.  The origin may be due to the possible antiferromagnetism of
nanographites in Sn-MG (see Sec.~\ref{disC}).

\begin{figure}
     \begin{center}
     \includegraphics*[width=8cm]{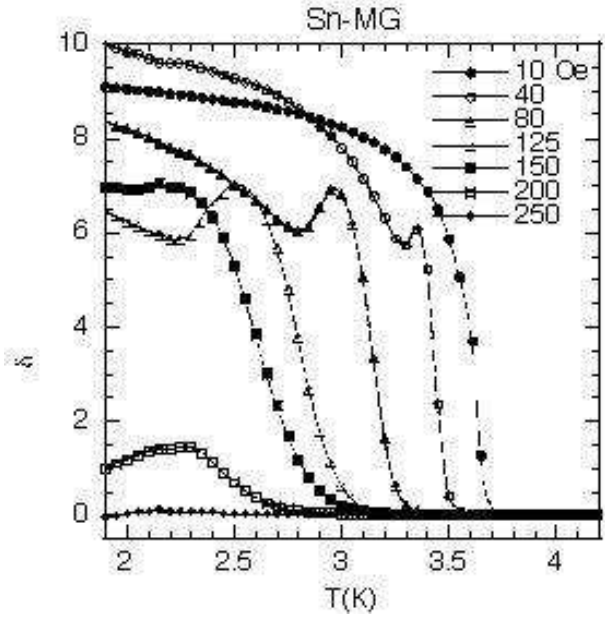}
     \end{center}
\caption{$T$ dependence of the difference $\delta$ (= $\chi_{FC} -
\chi_{ZFC}$) at low $H$ for Sn-MG. The solid lines are guides to the eyes.}
\label{fig:three}
\end{figure}

Figure \ref{fig:three} shows the $T$ dependence of the difference $\delta$
(= $\chi_{FC} - \chi_{ZFC}$) at various $H$.  The difference $\delta$ for
$H < H^{*}$ ($H^{*}\approx 40$ Oe) decreases with increasing $T$ and
reduces to zero at a characteristic temperature.  In contrast, the
difference $\delta$ for $H^{*} < H < 150$ Oe decreases in two steps with
increasing $T$, showing a local minimum and a local maximum.  The
difference $\delta$ shows a small peak at 2.25 K for $H = 200$ Oe and is
almost equal to zero for any $T$ for $H = 250$ Oe.

\subsection{\label{resultB}$M_{IR}$ and $M_{TR}$ }

\begin{figure}
     \begin{center}
     \includegraphics*[width=8cm]{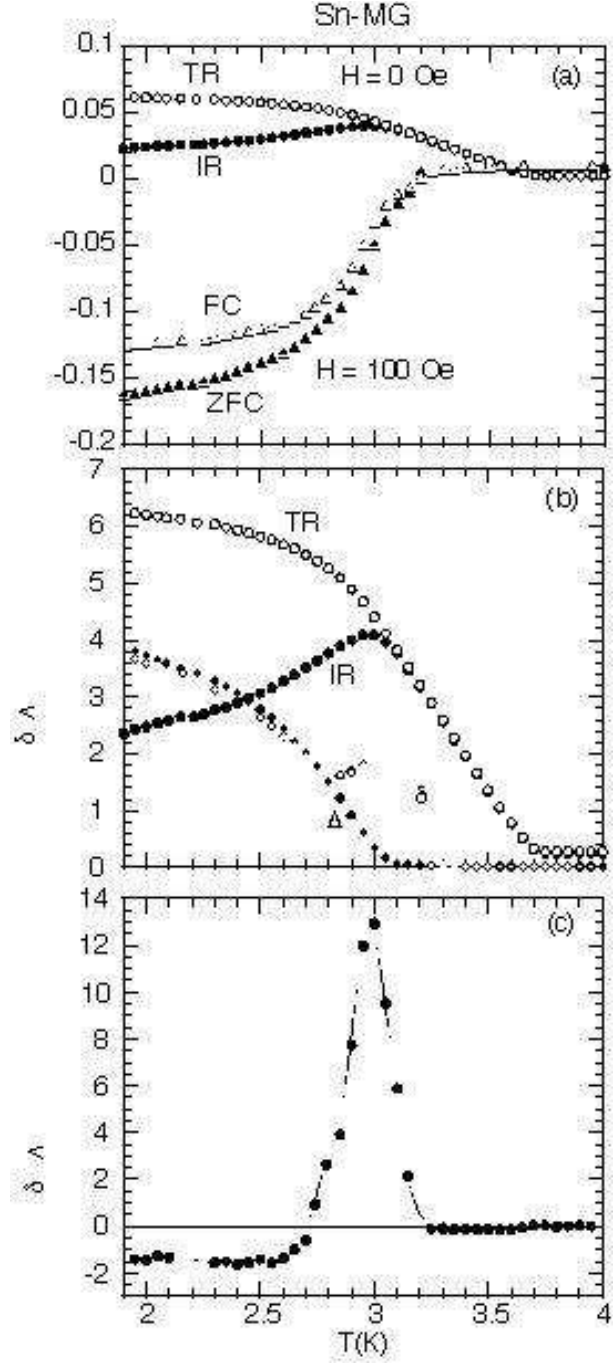}
     \end{center}
\caption{(a) $T$ dependence of $M_{ZFC}$, $M_{IR}$, $M_{FC}$, and $M_{TR}$
for Sn-MG. The definition of these magnetizations and the method of the
measurement are described in the text.  $H$ = 100 Oe.  (b) $T$ dependence
of $\delta$ (= $M_{FC} - M_{ZFC}$), $\Delta$ (= $M_{TR} - M_{IR}$),
$M_{IR}$, and $M_{TR}$.  (c) $T$ dependence of ($\delta - \Delta$).  The
solid line is guide to the eyes.}
\label{fig:four}
\end{figure}

In a mixed state of conventional type-II superconductors that have
well-defined lower critical field $H_{c1}$ and upper critical field
$H_{c2}$ ($H_{c1} < H < H_{c2}$), the thermodynamically most stable state
involves filling the system with fluxoids, exhibiting only a partial
Meissner effect with the magnetic induction $B = H + 4\pi M > 0$.  Even a
small degree of inhomogenity is sufficient to pin some fluxoids at local
free energy minima, so that some fluxoids remain trapped in place where $H$
is removed.  In Sn-MG the values of $M$ and $B$ are strongly dependent on
the magnetic states such as the ZFC, FC, IR (isothermal remnant), and TR
(thermoremnant) states.  Experimentally, the values of $M$ and $B$ for
these states are obtained as follows.  First the sample was cooled from 298
to 1.9 K at $H$ = 0.  Then $H$ was applied.  The measurements of $M_{ZFC}$
and $M_{IR}$ were done with increasing $T$ from 1.9 to 4 K. At each $T$,
$M_{ZFC}$ was measured at $H$ and then $M_{IR}$ was measured 100 sec later
after the field was changed from $H$ to 0 Oe.  Second, the sample was
annealed at 20 K for 1200 sec at $H$.  The measurements of $M_{FC}$ and
$M_{TR}$ were done with decreasing $T$ from 4 to 1.9 K. At each $T$,
$M_{FC}$ was measured at $H$ =100 Oe and then $M_{TR}$ was measured 100 sec
later after the field was changed from $H$ to 0 Oe.  The values of $M$ and
$B$ for each state thus obtained are described by $B_{ZFC} = H + 4\pi
M_{ZFC}$, $B_{IR} = 4\pi M_{IR}$, $B_{FC} = H + 4\pi M_{FC}$, and $B_{TR} =
4\pi M_{TR}$.  Here $B_{FC}$ and $B_{ZFC}$ are the magnetic inductions in
the FC and ZFC states with $H$, while $B_{IR}$ and $B_{TR}$ are the
magnetic inductions in the IR and TR states with $H = 0$, respectively. 
These magnetic inductions result from the flux trapping at pinning centers. 
Figure 4(a) shows the $T$ dependence of $M_{ZFC}$ and $M_{FC}$ at $H = 100$
Oe ($> H^{*}$), and $M_{IR}$ and $M_{TR}$ at $H = 0$.  The $T$ dependence
of $M_{IR}$ and $M_{TR}$ is different from $B_{IR}$ and $B_{TR}$ by only a
factor of 4$\pi$.  The $T$ dependence of $M_{IR}$ and $M_{TR}$ is similar
to that of typical spin glass systems such as 3D Ising spin glass
Fe$_{c}$Mn$_{1-c}$TiO$_{3}$ \cite{Gunnarsson1988}, where the competition
between the nearest neighbor antiferromagnetic and ferromagnetic
interactions leads to the spin frustration effect.  The magnetization
$M_{IR}$ exhibits a broad peak around 3.0 K, while $M_{TR}$ decreases with
increasing $T$.  Both $M_{IR}$ and $M_{TR}$ reduce to zero at $T_{c}$,
implying the appearance of both $B_{IR}$ and $B_{TR}$ below $T_{c}$.  The
deviation of $M_{IR}$ from $M_{TR}$ occurs below 3.1 K, showing the
irreversible effect of magnetization.  Note that the point at $T$ = 2.94 K
and $H$ = 100 Oe is located on the line $H_{gl}$ in the $H$-$T$ diagram
(see Sec.~\ref{disA}).  Figure \ref{fig:four}(b) shows the $T$ dependence
of $\Delta$ [=$M_{TR} - M_{IR} = (B_{TR} - B_{IR})/4\pi$] and $\delta$ [=
$M_{FC} - M_{ZFC} = (B_{FC} - B_{ZFC})/4\pi$], as well as $M_{IR}$ and
$M_{TR}$.  Figure \ref{fig:four}(c) shows the $T$ dependence of ($\delta -
\Delta$).  The difference ($\delta - \Delta$) is equal to zero at $T$ above
3.25 K and exhibits a peak at 3.0 K where $M_{IR}$ has a peak.  It becomes
negative below 2.7 K. The temperature region between 3.0 and 2.4 - 2.7 K
corresponds to the vortex glass phase for $H = 100$ Oe.

\begin{figure}
     \begin{center}
     \includegraphics*[width=8cm]{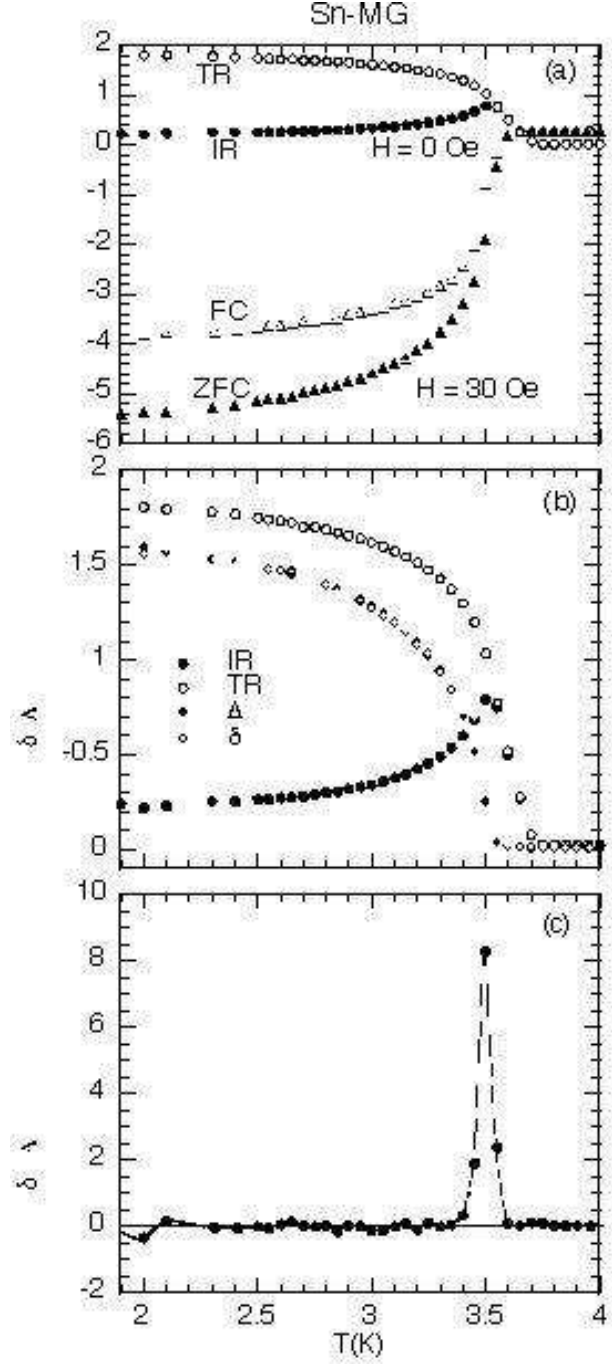}
     \end{center}
\caption{(a) $T$ dependence of $M_{ZFC}$, $M_{IR}$, $M_{FC}$, and $M_{TR}$
for Sn-MG. $H$ = 30 Oe.  (b) $T$ dependence of $\delta$ (= $M_{FC} -
M_{ZFC}$), $\Delta$ (= $M_{TR} - M_{IR}$), $M_{IR}$, and $M_{TR}$. 
(c) $T$ dependence of ($\delta - \Delta$).  The solid line is guide to the
eyes.}
\label{fig:five}
\end{figure}

Figures \ref{fig:five}(a) - (c) show the $T$ dependence of $M_{FC}$,
$M_{ZFC}$, $M_{IR}$, $M_{TR}$, $\delta$, $\Delta$, and $\delta - \Delta$
for $H$ = 30 Oe.  The magnetization $M_{IR}$ appears below $T_{c}$,
exhibiting a peak at 3.5 K. Note that a point at $T$ = 3.5 K and $H$ = 30
Oe is located on the line $H_{al}$ in the $H$-$T$ diagram (see
Sec.~\ref{disA}).  The deviation of $M_{IR}$ from $M_{TR}$ occurs at $T$
below 3.6 K, while the deviation of $M_{ZFC}$ from $M_{FC}$ occurs below
3.55 K. The difference ($\delta - \Delta$) exhibits a very sharp peak at
3.5 K, being equal to zero above 3.6 K and below 3.4 K.

\subsection{\label{resultC}$M$ vs $H$} 

\begin{figure}
     \begin{center}
     \includegraphics*[width=8cm]{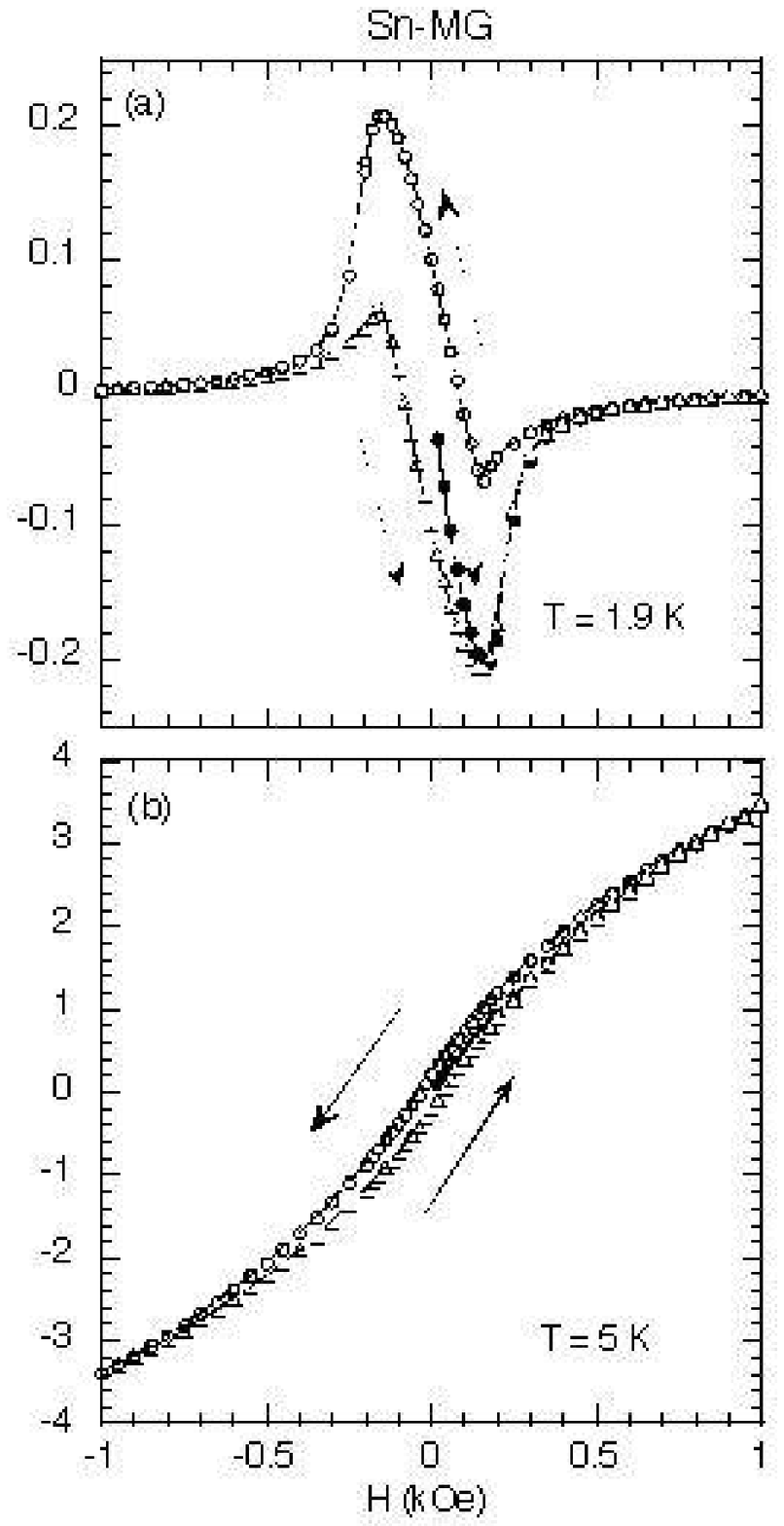}
     \end{center}
\caption{Magnetization loop ($M$ vs $H$) for Sn-MG. (a) $T$ = 1.9 K and (b)
5 K. The method of the measurement is described in the text.  The solid
lines are guides to the eyes.}
\label{fig:six}
\end{figure}

Figure \ref{fig:six} shows the hysteresis loop of the magnetization $M$ at
$T = 1.9$ and 5 K. The $M$-$H$ curve at 1.9 K shows a typical
superconducting behavior with a negative local minimum around $H = 160$ Oe,
corresponding to the phase boundary (the line $H_{ag}$) between the vortex
lattice and vortex glass phase (see Sec.~\ref{disA}).  The $M$-$H$ curve is not
reversible, partly because some fluxoids remain trapped at pinning centers
when $H$ is removed.  The $M$-$H$ curve at 5 K is not perfectly
proportional to $H$.  The susceptibility $M/H$ is estimated to be $3.45
\times 10^{-6}$ emu/g at $H = 1$ kOe, which is in contrast to the negative
diamagnetic susceptibility of pristine graphite: $\chi = - 30 \times
10^{-6}$ emu/g for $H$ along the $c$ axis and $-5.5 \times 10^{-7}$ emu/g
along the basal plane (so-called $c$ plane) perpendicular to the $c$ axis
below 100 K \cite{Heremans1994}.

\begin{figure}
     \begin{center}
     \includegraphics*[width=8cm]{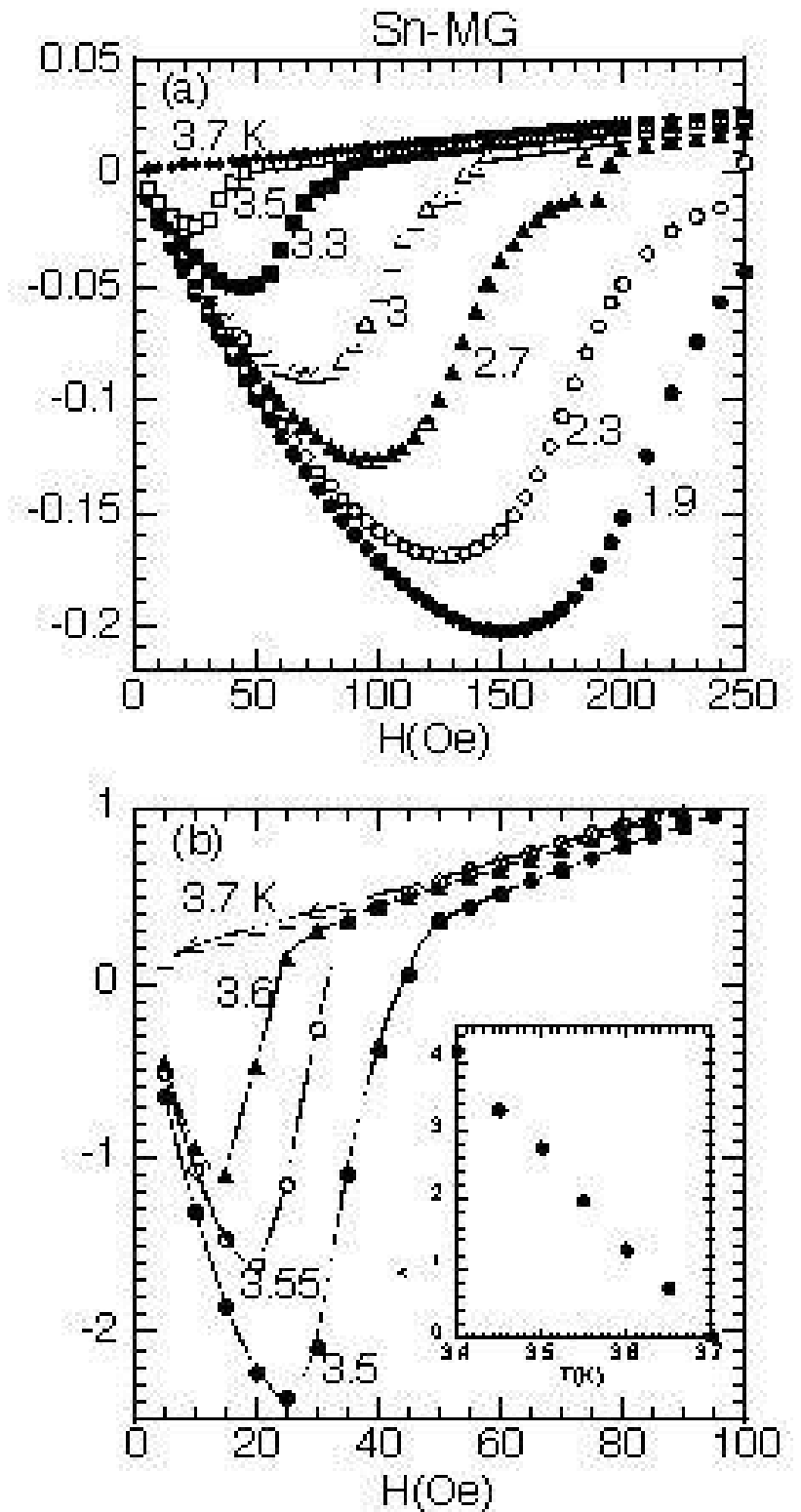}
     \end{center}
\caption{(a) and (b) $M_{ZFC}$ vs $H$ at various $T$.  The solid lines are
guides to the eyes.  The inset of Fig.~\ref{fig:seven}(b) shows the jump in
magnetization $\Delta M_{ZFC}$ (which is defined in the text) as a
function of $T$.}
\label{fig:seven}
\end{figure}

Figures ~\ref{fig:seven}(a) and (b) show typical data of $M_{ZFC}$ vs $H$
at various $T$.  The sample was cooled from 298 K to $T$ ($< 4$ K) at $H =
0$.  The magnetization $M_{ZFC}$ at $T$ was measured with increasing $H$
($0 < H < 250$ Oe).  The magnetization $M_{ZFC}$ exhibits a single local
minimum at a characteristic field for $T < T_{c}$.  Similar behavior has
been reported in YBa$_{2}$Cu$_{3}$O$_{6.92}$ by Shibata et al. 
\cite{Shibata2002}.  It seems that a discontinuous change in $M_{ZFC}$ is
observed just above the characteristic field for $T^{*} < T < T_{c}$, where
$T^{*} = 3.4$ K and $T_{c} = 3.75$ K. The field of the local-minimum for
$M_{ZFC}$ vs $H$, which is plotted as a function of $T$, forms the line
$H_{al}$ of the first order transition in the $H$-$T$ plane (see
Sec.~\ref{disA}).  The jump height $\Delta M_{ZFC}$, which is defined as
the difference between the value of $M_{ZFC}$ at which $M_{ZFC}$ starts to
be proportional to $H$ and the local-minimum value of $M_{ZFC}$ vs $H$,
rapidly decreases with increasing $T$ and reduces to zero at $T_{c}$.  The
jump height $\Delta M_{ZFC}$ may be described by $\Delta M_{ZFC} =M_{l} -
M_{a}$ \cite{Welp1996}, where $M_{l}$ and $M_{a}$ denote the magnetization
in the vortex liquid and the vortex lattice phase, respectively.  The
positive value of $\Delta M_{ZFC}$ implies that the vortex density in the
vortex liquid phase is higher than that in the vortex lattice phase.  The
magnetization $M_{ZFC}$ at $T = 3.7$ and 3.8 K is almost proportional to
$H$ with a positive slope.  The derivative d$M_{ZFC}$/d$H$ at 3.7 and 3.8 K
shows a broad peak around $H = 100$ Oe, suggesting the existence of AF
short-range order.  The origin of the AF short-range order is due to
nanographites in Sn-MG (see Sec.~\ref{disC}).  Note that a discontinuous
jump of $M_{ZFC}$ in the vicinity of $M_{ZFC} = 0$ is an artifact due to
the SQUID measurement.  Such a jump in the magnetization is frequently
observed when the magnetization changes its sign.

The lower critical field $H_{c1}(T)$ is defined as a field at which 
the curve of $M$ vs $H$ starts to deviate from the linear portion due 
to the penetration of the magnetic flux into the system: $H_{c1}(T = 
1.9$ K$) = 20 \pm 5$ Oe. 

\subsection{\label{resultD}$\chi^{\prime}$ and $\chi^{\prime\prime}$}

\begin{figure}
     \begin{center}
     \includegraphics*[width=8cm]{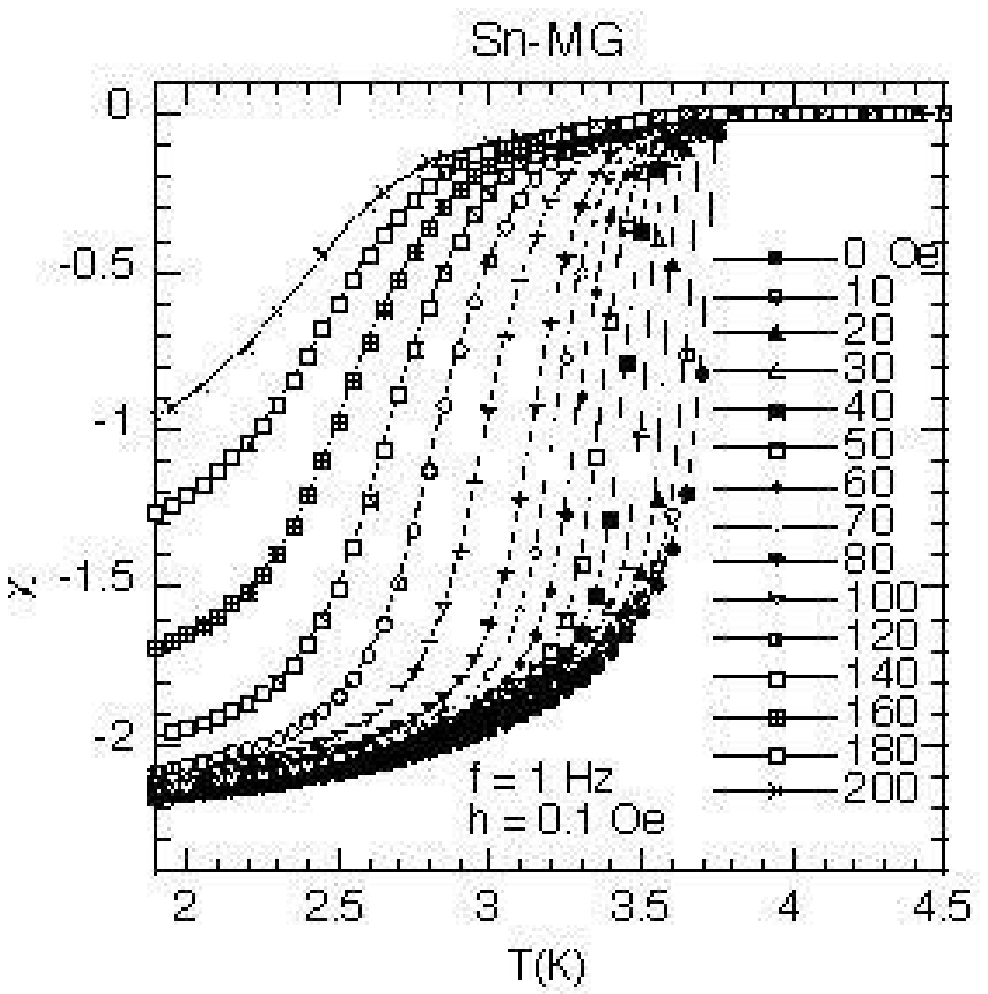}
     \end{center}
\caption{$T$ dependence of $\chi^{\prime}$ at various $H$ for Sn-MG. $h =
0.1$ Oe.  $f = 1$ Hz.  The solid lines are guides to the eyes.}
\label{fig:eight}
\end{figure}

Figure \ref{fig:eight} shows the $T$ dependence of the dispersion
$\chi^{\prime}$ at various $H$, respectively, where $f = 1$ Hz and $h =
0.1$ Oe.  The AC response to such a small $h$ is found to be linear in the
whole range of temperatures and fields.  The derivative
d$\chi^{\prime}$/d$T$ shows a sharp peak at $T_{c}$ at $H$ = 0, shifting to
the low-$T$ side with increasing $H$.  The peak temperature for
d$\chi^{\prime}$/d$T$ vs $T$ plotted as a function of $H$ in the $H$-$T$
diagram forms the line $H_{gl}$ for $H > H^{*}$ and the line $H_{al}$ for
$H < H^{*}$ (see Sec.~\ref{disA}).

\begin{figure}
     \begin{center}
     \includegraphics*[width=8cm]{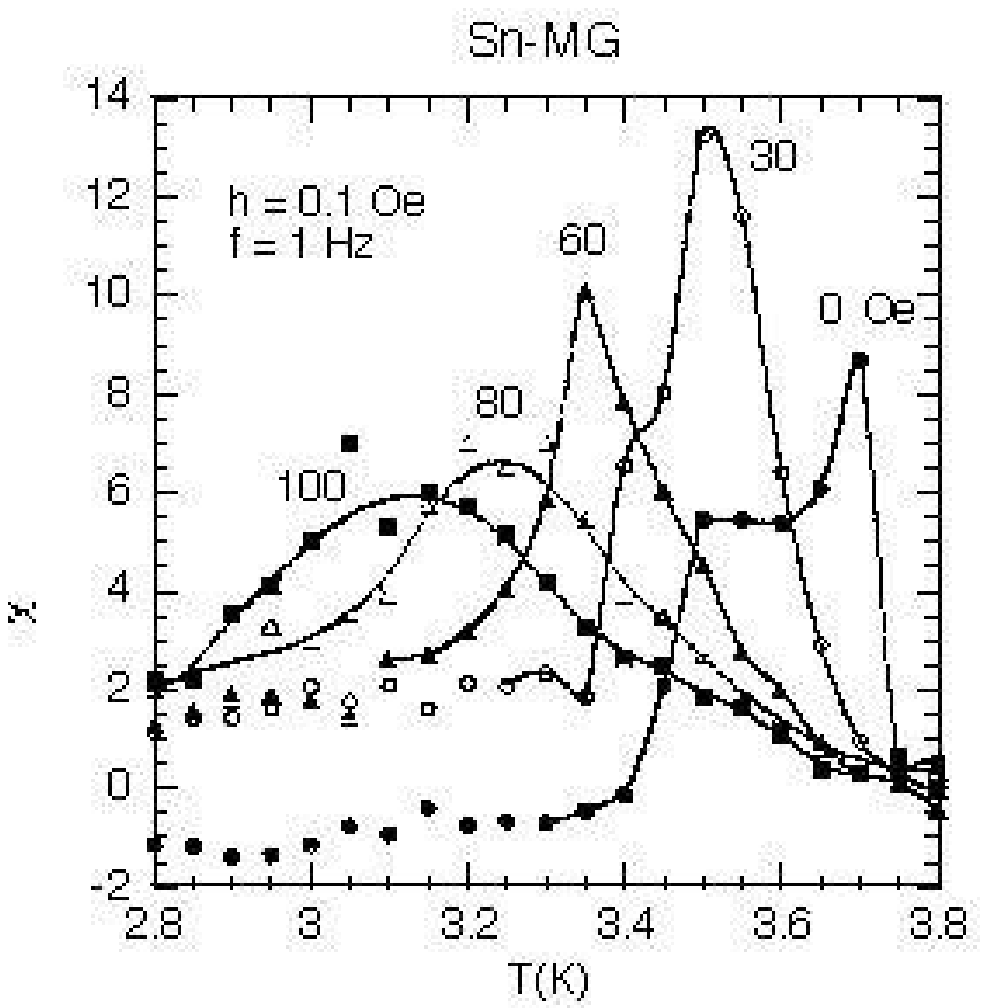}
     \end{center}
\caption{$T$ dependence of $\chi^{\prime\prime}$ at various $H$ for Sn-MG.
$h$ = 0.1 Oe.  $f = 1$ Hz.  The solid lines are guides to the eyes.}
\label{fig:nine}
\end{figure}

Figure \ref{fig:nine} shows the $T$ dependence of the absorption
$\chi^{\prime\prime}$ at various $H$, where $f = 1$ Hz and $h = 0.1$ Oe. 
The absorption $\chi^{\prime\prime}$ at $H = 0$ seems to have two peaks at
$T_{c}$ and 3.5 K. The peak at $T_{c}$ at $H$ = 0 shifts to the low-$T$
side with increasing $H$.  In contrast, the peak at 3.5 K at $H = 0$
remains unshifted, becoming a shoulder even at 10 Oe, and tends to
disappear above $H = 30$ Oe.  Note that the peak temperature of the main
peak for $\chi^{\prime\prime}$ vs $T$ is slightly higher than that for
d$\chi^{\prime}$/d$T$ vs $T$ at the same $H$.  The peak temperature for
$\chi^{\prime\prime}$ vs $T$ is plotted as a function of $H$ in the $H$-$T$
diagram, forming the line $H_{al}$ for $H < H^{*}$ and a new line
$H_{gl^{\prime}}$ for $H > H^{*}$ which deviates from the line $H_{gl}$ for
$H^{*} < H < 150$ Oe (see Sec.~\ref{disA}).  The main peak of
$\chi^{\prime\prime}$ vs $T$ is sharp for $H < H^{*}$ but becomes much
broad for $H > H^{*}$.

\subsection{\label{resultE}$\Theta_{1}^{\prime}/h$ and
$\Theta_{1}^{\prime\prime}/h$}

\begin{figure}
     \begin{center}
     \includegraphics*[width=8cm]{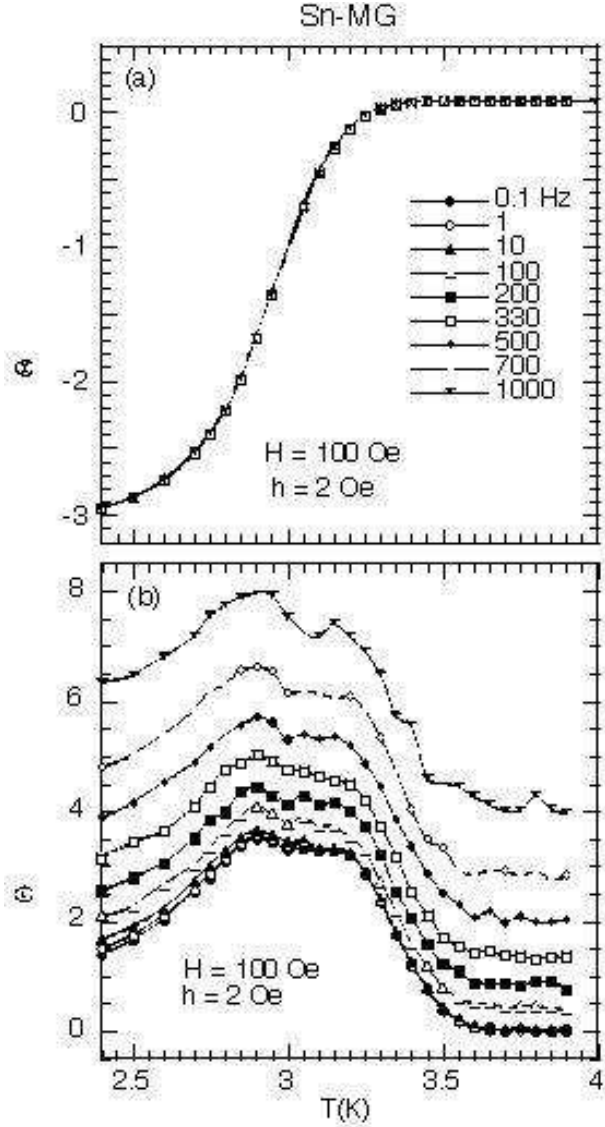}
     \end{center}
\caption{$T$ dependence of (a) the dispersion $\Theta_{1}^{\prime}/h$ and
(b) the absorption $\Theta_{1}^{\prime\prime}/h$ at various $f$ ($0.1 \leq
f \leq 1000$ Hz) for Sn-MG. $H = 100$ Oe.  Because of the nonlinearlity,
$\Theta_{1}^{\prime\prime}/h$ at $h = 2$ Oe is not equal to
$\chi^{\prime\prime}$ at $h = 0.1$ Oe (see Fig.~\ref{fig:twelve}(b) for the
nonlinearity at $H = 0$).  The solid lines are guides to the eyes.}
\label{fig:ten}
\end{figure}

The AC response ($\Theta_{1}^{\prime}/h$ and $\Theta_{1}^{\prime\prime}/h$)
for sufficiently large amplitude of the AC magnetic field becomes nonlinear
below $T_{c}$.  In fact, as it will be shown later, the $T$ dependence of
$\Theta_{1}^{\prime\prime}/h$ is strongly dependent on $h$ for $h \geq 0.2$
Oe.  This is in contrast to the case of $h = 0.1$ Oe, where the linearity
is valid: $\Theta_{1}^{\prime}/h = \chi^{\prime}$ and
$\Theta_{1}^{\prime\prime}/h = \chi^{\prime\prime}$.  Here a large
amplitude of the AC magnetic field ($h = 2$ Oe) was used to examine the
effect of nonlinearity on the vortex glass phase.  Figure \ref{fig:ten}
shows the $T$ dependence of $\Theta_{1}^{\prime}/h$ and
$\Theta_{1}^{\prime\prime}/h$ for Sn-MG at various $f$ ($0.1 \leq f \leq
1000$ Hz) in the presence of $H$ (= 100 Oe), where $h = 2$ Oe.  The $T$
dependence of $\Theta_{1}^{\prime}/h$ is almost independent of $f$, while
the $T$ dependence of $\Theta_{1}^{\prime\prime}/h$ is strongly dependent
on $f$.  The shape of $\Theta_{1}^{\prime\prime}/h$ vs $T$ does not change
with $f$ but the basement of $\Theta_{1}^{\prime\prime}/h$ proportionally
increases with increasing $f$ for $0.1 \leq f \leq 1000$ Hz at any $T$
below $T_{c}$.  As shown in Fig.~\ref{fig:ten}(b) there are two broad peaks
at 2.90 and 3.15 K for $\Theta_{1}^{\prime\prime}/h$ with $h = 2$ Oe. 
These peaks do not shift with $f$ for $0.1 \leq f \leq 1000$ Hz within the
experimental error in $T$ (typically $\Delta T_{\epsilon} = 0.01$ K).  No
relaxation time effect is observed, suggesting $\omega \tau \ll 1$, where
$\tau$ is the relaxation time of the vortex system and $\omega$ (=$ 2\pi
f$) is the measured angular frequency.  This is very different from the
case of typical spin-glass systems where $\omega \tau \approx 1$ and the
peak of $\Theta_{1}^{\prime\prime}/h$ shifts to the high $T$ side with
increasing $f$ \cite{Mydosh1993}.  We show that the $T$ and $f$ dependence
of $\Theta_{1}^{\prime}/h$ and $\Theta_{1}^{\prime\prime}/h$ are explained
in terms of the screening behavior arising from skin size effects.  The
absorption peak occurs when the electromagnetic penetration depth $\delta
_{s}$ becomes comparable to the sample size $L$ (see Sec.~\ref{disB}).  The
peak at 3.15 K coincides with that of $\chi^{\prime\prime}$ at $h = 0.1$ Oe
and $f = 1$ Hz (see Fig.~\ref{fig:nine}).  The point at $H = 100$ Oe and $T
= 3.15$ K is located on the line $H_{gl^{\prime}}$.  On the other hand, the
peak at 2.9 K is newly observed for $h = 2$ Oe.  The point at $H$ = 100 Oe
and $T = 2.9$ K is located on the line $H_{gl}$.  Note that the data of
d$\chi^{\prime}$/d$T$ vs $T$ shows a peak at 2.94 K and d$\delta$/d$T$ has
a local minimum at 3.0 K. In this sense, the line $H_{gl}$ is the
irreversibility line.

\begin{figure}
     \begin{center}
     \includegraphics*[width=8cm]{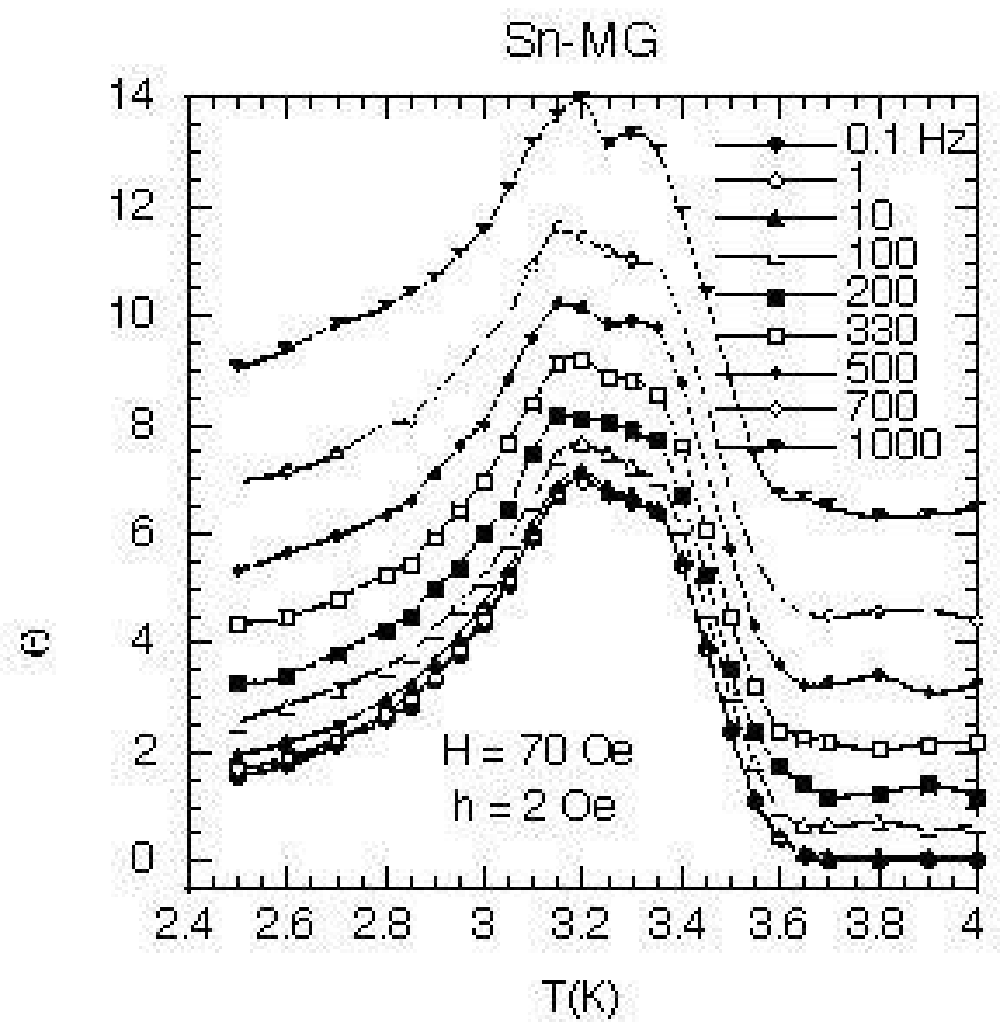}
     \end{center}
\caption{$T$ dependence of $\Theta_{1}^{\prime\prime}/h$ at various $f$
($0.1 \leq f \leq 1000$ Hz) for Sn-MG. $H = 75$ Oe.  $h = 2$ Oe.  The solid
lines are guides to the eyes.}
\label{fig:eleven}
\end{figure}

\begin{figure}
     \begin{center}
     \includegraphics*[width=8cm]{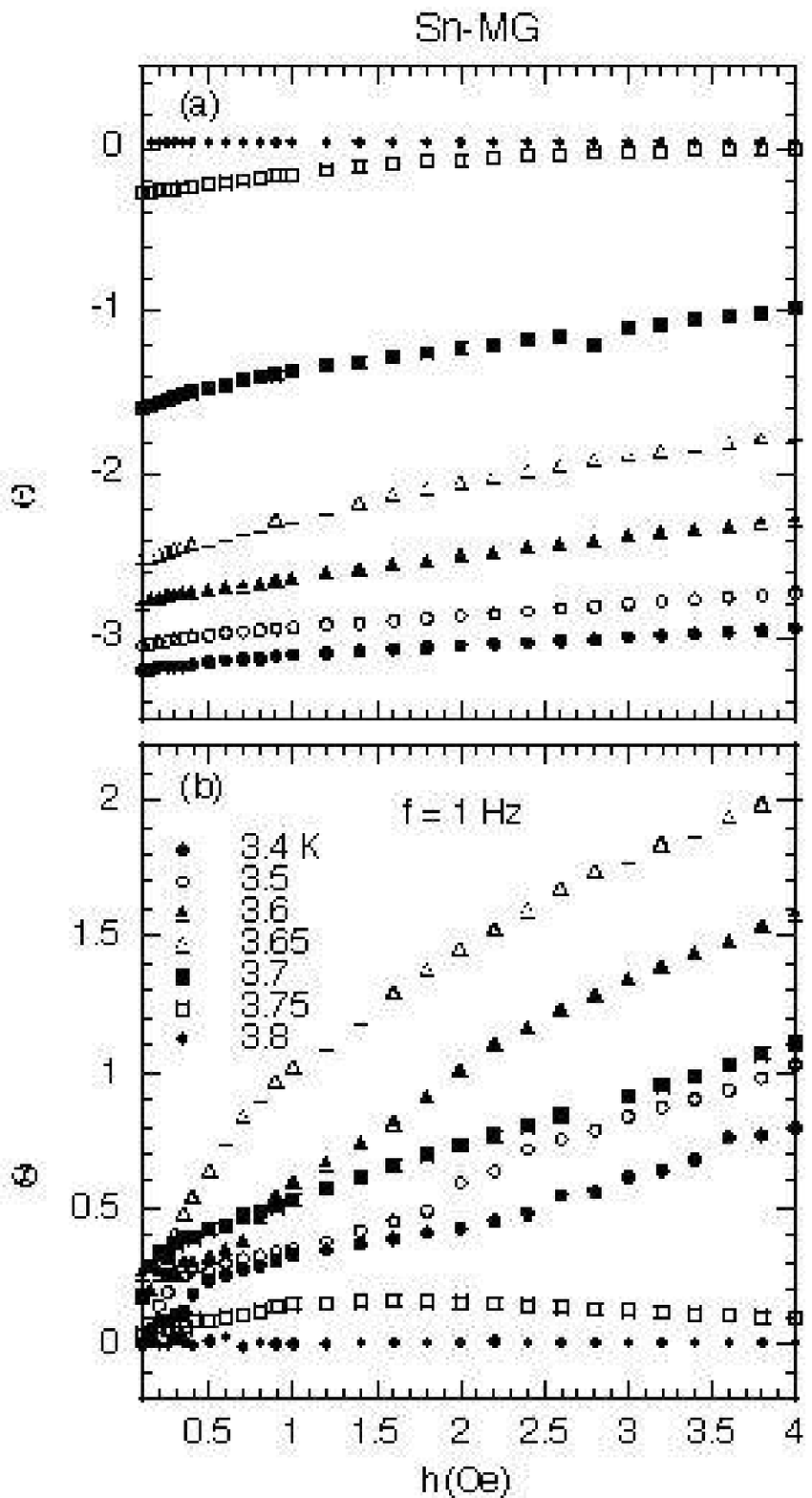}
     \end{center}
\caption{$h$ dependence of (a) $\Theta_{1}^{\prime}/h$ and (b)
$\Theta_{1}^{\prime\prime}/h$ at various $T$ for Sn-MG. $f = 1$ Hz.  $H =
0$.}
\label{fig:twelve}
\end{figure}

Figure \ref{fig:eleven} shows the $T$ dependence of
$\Theta_{1}^{\prime\prime}/h$ for Sn-MG at various $f$ ($0.1 \leq f \leq
1000$ Hz) in the presence of $H$ (= 70 Oe), where $h = 2$ Oe.  There are
two peaks at 3.1 - 3.15 K and 3.3 - 3.4 K. It seems that these peaks do not
shift with $f$.  This result suggests no occurrence of the relaxation time
effect for $0.1 \leq f \leq 1000$ Hz.  The peak at 3.3 - 3.4 K coincides
with that at 3.34 K of $\chi^{\prime\prime}$ at $h = 0.1$ Oe and $f = 1$ Hz
(the data at $H = 70$ Oe are not shown in Fig.  9).  The point at $H = 70$
Oe and $T = 3.3$ - 3.4 K is located on the line $H_{gl^{\prime}}$.  On the
other hand, the peak at 3.1 - 3.15 K is newly observed for $h = 2$ Oe.  The
point at $H = 70$ Oe and $T = 3.1$ - 3.15 K is located on the line
$H_{gl}$.  Figure \ref{fig:twelve} shows the $h$ dependence of (a)
$\Theta_{1}^{\prime}/h$ and (b) $\Theta_{1}^{\prime\prime}/h$ for Sn-MG at
various $T$ in the absence of $H$, where $0.01 \leq h \leq 4$ Oe.  Both
$\Theta_{1}^{\prime}/h$ and $\Theta_{1}^{\prime\prime}/h$ are not
independent of $h$ for $h > 0.2$ Oe, indicating the nonlinearity of the AC
response.  The dispersion $\Theta_{1}^{\prime}/h$ linearly changes with $h$
below $T_{c}$, but is almost independent of $h$ at 3.8 K just above
$T_{c}$.

\section{\label{dis}Discussion}
\subsection{\label{disA}$H$-$T$ phase diagram}

\begin{figure}
     \begin{center}
     \includegraphics*[width=8cm]{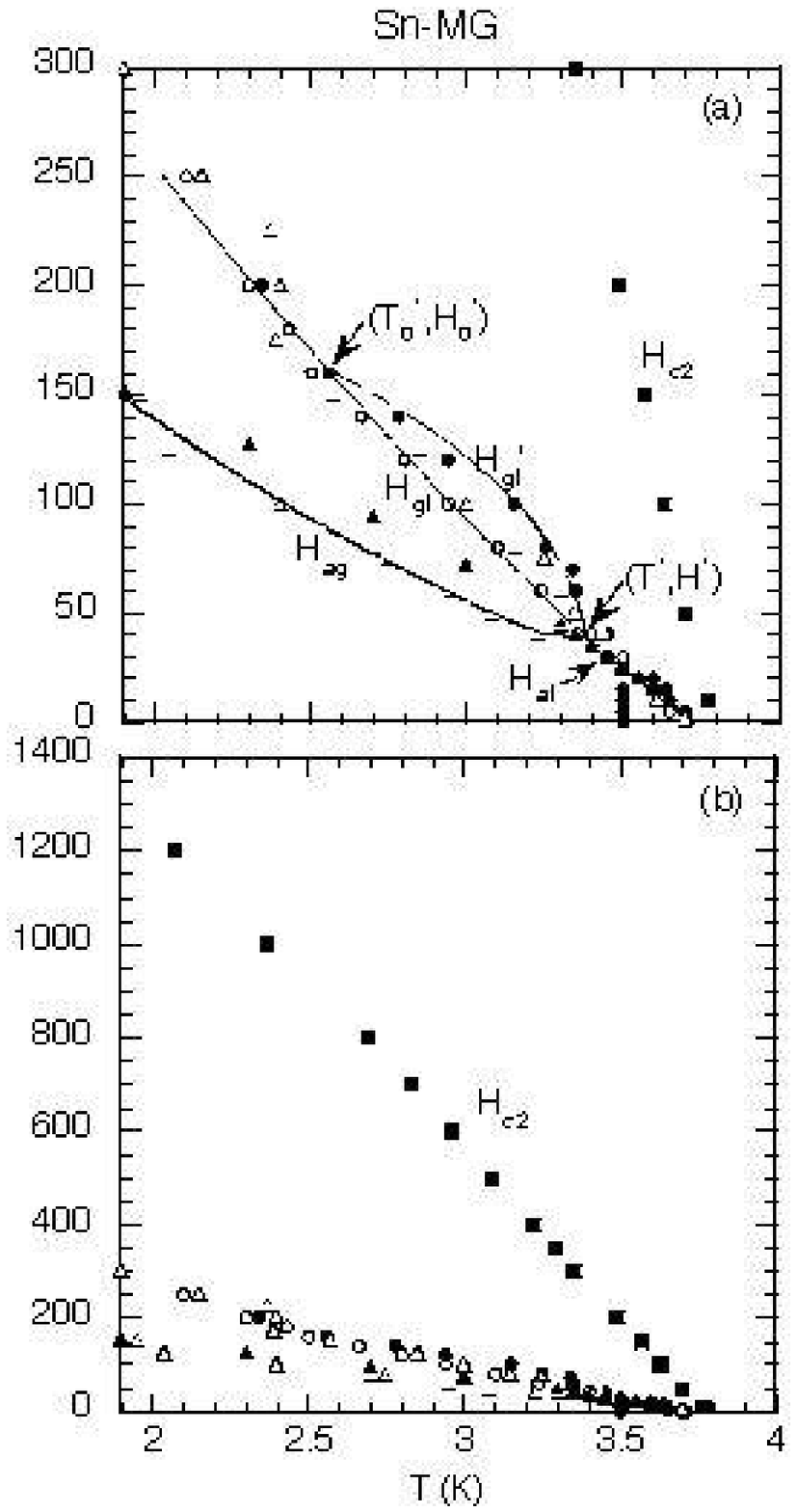}
     \end{center}
\caption{(a) and (b) $H$-$T$ diagram for Sn-MG. The peak temperatures of
d$\chi^{\prime}$/d$T$ vs $T$ ({\Large $\circ$}), $\chi^{\prime\prime}$ vs
$T$ ({\Large $\bullet$}), and the local-minimum temperatures of
d$\delta$/d$T$ vs $T$ ($\triangle$) are plotted as a function of $H$.  The
local-minimum fields of $M_{ZFC}$ vs $H$ ($\blacktriangle$) are plotted as
a function of $T$.  The vortex lattice (Abrikosov lattice), vortex liquid,
and vortex glass phases are separated by the lines $H_{al}$, $H_{ag}$,
$H_{gl}$, and $H_{gl^{\prime}}$.  These lines merge at the multicritical
point located at ($T^{*} \approx 3.4$ K, $H^{*} \approx 40$ Oe).  The line
$H_{c2}$ is a crossover line denoted by closed squares ($\blacksquare$). 
The lower critical line $H_{c1}$ is not shown in this figure.  $H_{c2}(T =
0$K$) \approx 1.85$ kOe.  The solid lines for the lines $H_{gl}$,
$H_{gl^{\prime}}$, $H_{ag}$, and $H_{al}$ denote the least-squares fitting
curves of the data for $H_{i}$ vs $T$ to Eq.(\ref{eq:one}) with $i = gl,
gl^{\prime}, ag,$ and $al$, respectively.  The corresponding parameters are
given in the text.}
\label{fig:thirteen}
\end{figure}

Our results obtained above are summarized in the $H$-$T$ diagram.  In
Figs.~\ref{fig:thirteen}(a) and (b), we make a plot of the peak
temperatures of d$\chi^{\prime}$/d$T$ vs $T$ and $\chi^{\prime\prime}$ vs
$T$, and the local-minimum temperatures of d$\delta$/d$T$ vs $T$ as a
function of $H$, and the local-minimum fields of $M_{ZFC}$ vs $H$ as a
function of $T$.  We find that these data lie on the phase boundaries
denoted by the lines $H_{ag}$, $H_{gl}$, $H_{gl^{\prime}}$, and $H_{al}$,
which separate the vortex lattice, vortex glass, and vortex liquid phase. 
These lines merge at the multicritical point at $T^{*} = 3.4$ K and $H^{*}
= 40$ Oe.  The lines $H_{ag}$, $H_{gl}$, and $H_{gl^{\prime}}$ are of the
second order, while the line $H_{al}$ is of the first order.  The line
$H_{c2}$ is a crossover line separating the normal phase at higher
temperatures from the vortex liquid phase.  (i) On the line $H_{gl}$, the
data of d$\chi^{\prime}$/d$T$ vs $T$ have a peak at fixed $H$, the data of
d$\delta$/d$T$ vs $T$ have a local minimum, and the data of
$\chi^{\prime\prime}$ vs $T$ have a peak for $H \geq 160$ Oe.  (ii) On the
line $H_{gl^{\prime}}$, the data of $\chi^{\prime\prime}$ vs $T$ have a
peak at fixed $H$.  (iii) On the line $H_{ag}$, the data of d$\delta$/d$T$
vs $T$ have a local minimum at fixed $H$ and the data of $M_{ZFC}$ vs $H$
have a local minimum at fixed $T$.  (iv) On the line $H_{al}$, the data of
d$\chi^{\prime}$/d$T$ vs $T$ and $\chi^{\prime\prime}$ vs $T$ have a peak
at fixed $H$, the data of d$\delta$/d$T$ vs $T$ have a local minimum at
fixed $H$, and the data of $M_{ZFC}$ vs $H$ have a local minimum at fixed
$T$.  (v) On the line $H_{c2}$, the data of $\chi_{ZFC}$ vs $T$ come to
take a saturated value $\chi_{ZFC}^{(s)}$ at fixed $H$.  The susceptibility
$\chi_{ZFC}^{(s)}$ is positive and weakly dependent on $H$.  It takes $1.8
\times 10^{-7}$ emu/g at $H = 10$ Oe, increasing with increasing $H$, and
reaches a peak ($= 4.63 \times 10^{-6}$ emu/g) at $H = 200$ Oe.  In turn it
decreases with further increasing $H$, and reaches $3.75 \times 10^{-6}$
emu/g at $H = 1$ kOe (see Fig.~\ref{fig:three}(a)).

Using an empirical relation given by Werhammer et al. 
\cite{Werthamer1966}, $H_{c2}(T = 0$ K) $= - 0.69
T_{c}($d$H_{c2}$/d$T)_{T_{c}}$, the value of $H_{c2}$ at $T = 0$ K can be
estimated as $H_{c2}(0) = 1850 \pm 20$ Oe, where we use $T_{c} = 3.75$ K
and a slope (d$H_{c2}$/d$T$)$_{T_{c}} = - 714.4 \pm 6.8$ (Oe/K) obtained
from the linear relation in $H_{c2}$ vs $T$ in the vicinity of $T = T_{c}$
and $H_{c2} = 0$.  The coherence length $\xi (0)$ can be estimated as $420
\pm 20 \AA$ using a relation $H_{c2}(0) = \Phi_{0}/(2\pi \xi (0)^{2})$,
where $\Phi_{0} (= 2.0678 \times 10^{-7}$ Gauss cm$^{2}$) is a fluxoid. 
Note that our value of $H_{c2}$(0) for Sn-MG is on the same order as that
of Sn films grown on an InSb (110) surface \cite{Didschuns2002}.  The
value of $H_{c2}(0)$ strongly depends on the thickness $t$ of the Sn films:
$H_{c2}(0) = 3160$ Oe for $t = 420 \AA$ and 2540 Oe for $t = 2540 \AA$ for
$H$ parallel to the Sn film.

Here we assume that the $T$ dependence of the line $H_{i}$ ($i = gl,
gl^{\prime}, ag,$ and $al$) may be described by a power law form given by
\begin{equation} 
    H = H_{i}^{*}(1-T/T_{i}^{*})^{\alpha (i)}, 
    \label{eq:one}
\end{equation} 
where $\alpha (i)$ is an exponent, and $H_{i}^{*}$ and
$T_{i}^{*}$ are characteristic field and temperature, respectively.  The
least squares fit of the data of $H$ vs $T$ ($40 \leq H \leq 250$ Oe)
obtained from the peak temperature of d$\chi^{\prime}$/d$T$ vs $T$ at $H$,
to Eq.(\ref{eq:one}) with $i = gl$ yields the parameters $\alpha (gl) =
1.43 \pm 0.05$, $T_{gl}^{*} = 3.90 \pm 0.08$ K, and $H_{gl}^{*} = 730 \pm
20$ Oe.  The exponent $\alpha (gl)$ is close to that (= 1.50) predicted
from the AT theory \cite{Almeida1978}, suggesting that the line $H_{gl}$ is
the so-called irreversibility line separating a higher-temperature vortex
liquid phase from a lower-temperature vortex glass phase.  The least
squares fit of the data of $H$ vs $T$ ($70 \leq H \leq 160$ Oe) obtained
from the peak temperature of $\chi^{\prime\prime}$ vs $T$ at $H$, to
Eq.(\ref{eq:one}) with $i = gl^{\prime}$ yields the parameters $\alpha
(gl^{\prime}) = 0.57 \pm 0.10$, $T_{gl^{\prime}}^{*} = 3.57 \pm 0.09$ K,
and $H_{gl^{\prime}}^{*} = 330 \pm 10$ Oe.  The exponent $\alpha
(gl^{\prime})$ is close to that (= 0.50) predicted from the Gabay-Thoules
(GT) theory \cite{Gabay1981}.  Such an occurrence of the AT-GT crossover
has been reported in granular high $T_{c}$ superconductors
\cite{Vieira2003,Vieira2001}.  No GT-power law form has been observed in a
single crystal of YBa$_{2}$Cu$_{3}$O$_{7}$ without granularity
\cite{Vieira2003}.  These results suggest that the AT-GT crossover behavior
is indicative of the frustrated nature of the system due to disorder and
granularity.  This crossover is an analogy to the Ising-Heisenberg
crossover behavior observed in Heisenberg-like spin glasses
\cite{Kenning1991}.  Such a crossover occurs when the magnetic field
collapses the random local anisotropy field.  The AT and GT lines in 
SG systems represent
the longitudinal and transverse freezing, respectively.

Similarly, the least-squares fit of the data of $H$ vs $T$ ($40 \leq H \leq
150$ Oe) obtained from the peak temperature of d$\chi^{\prime}$/d$T$ vs $T$
at $H$, to Eq.(\ref{eq:one}) with $i = ag$ yields the parameters $\alpha
(ag) = 2.06 \pm 0.05$, $T_{ag}^{*} = 4.77 \pm 0.80$ K, and $H_{ag}^{*} =
435 \pm 50$ Oe.  Note that the lower critical field $H_{c1}(0)$ is 
much lower than $H_{ag}^{*}$.  The least-squares fit of the data of $H$ vs $T$ ($0
\leq H \leq 40$ Oe) obtained from the peak temperature of
d$\chi^{\prime}$/d$T$ vs $T$ at $H$, to Eq.(\ref{eq:one}) with $i = al$
yields the parameters $\alpha (al) = 1.02 \pm 0.03$, $T_{al}^{*} = 3.70 \pm
0.02$ K, and $H_{al}^{*} = 580 \pm 50$ Oe.  The value of $\alpha (al)$ is
in good agreement with the mean-field value (= 1) derived from the
Ginzburg-Landau theory \cite{Ketterson1998}, where the correlation length
varies with $(1-T/T_{al}^{*})^{-1/2}$.

The $H$-$T$ diagram of Sn-MG is similar to that of a quasi-2D
superconductors (type-II).  The field for the multicritical point ($=
H^{*}$) is described by the dimensional crossover field that separates the
region of 2D thermal fluctuations (for $H > H^{*}$) and the 3D thermal
fluctuations ($H < H^{*}$).  In this sense, the $H$-$T$ diagram may be
universal for any quasi-2D superconductors when the $H$ axis and $T$ axis
are scaled by $H/H^{*}$ and $T/T_{c}$, respectively.  For
YBa$_{2}$Cu$_{3}$O$_{7}$ with $T_{c} = 90$ K, the multicritical point is
located at $T^{*} = 76$ K and $H^{*} = 90$ kOe \cite{Gammel1998}.  The
ratio $T^{*}/T_{c}$ for YBa$_{2}$Cu$_{3}$O$_{7}$ is 0.84, which is almost
equal to that (= 0.91) for Sn-MG with $T^{*} = 3.4$ K and $T_{c} = 3.75$ K.

\subsection{\label{disB}Nature of the lines $H_{gl}$ and $H_{gl^{\prime}}$}
Here we discuss the origin of the irreversibility lines $H_{gl^{\prime}}$
and $H_{gl}$ for $H^{*} < H < 160$ Oe.  We find the following results on
the $T$ dependence of AC magnetic susceptibility at small AC field ($h =
0.1$ Oe) and large AC field ($h = 2$ Oe).  (i) The linear absorption
$\chi^{\prime\prime}$ at $h = 0.1$ Oe exhibits a peak only on the line
$H_{gl^{\prime}}$.  (ii) The nonlinear absorption
$\Theta_{1}^{\prime\prime}/h$ exhibits a peak on the line $H_{gl^{\prime}}$
and a peak on the line $H_{gl}$.

It is assumed that there is an irreversibility line $T_{gl}(H)$ (or the
line $H_{gl}(T)$) in the $H$-$T$ plane which separates an Ohmic region [$T
> T_{gl}(H)$] from an non-Ohmic region [$T < T_{gl}(H)$].  Above
$T_{gl}(H)$, a voltage $V$ is proportional to current $I$, whereas below
$T_{gl}(H)$, $V$ shows an extremely nonlinear dependence where the linear
resistivity is zero and the electric field $E$ is described by $E \approx
\exp (-A/J^{\mu})$ \cite{Geshkenbein1991}.  Here $A$ and $\mu$ are positive
constants and $J$ is the current density.

Why does the peak of the absorption for small AC field amplitude appear 
on only  
the line $H_{gl^{\prime}}$ for $H > H^{*}$?  Why does the peak of the
absorption for large AC field amplitude appear on both the lines $H_{gl}$ and
$H_{gl^{\prime}}$ for $H > H^{*}$?  Similar problem was addressed first by
Geshkenbein et al.  \cite{Geshkenbein1991} and later by Steel and Graybeal
\cite{Steel1992}.  According to their skin size effect hypothesis, the peak
of $\Theta_{1}^{\prime\prime}/h$ appears when the electromagnetic
penetration depth $\delta_{s} = (c^{2} \rho /f)^{1/2}/2\pi$ is on the order
of the system size $L$, or when $\omega$ coincides with the characteristic
angular frequency $\omega_{peak}$ given by
\begin{equation} 
    \omega_{peak} = 0.8c^{2}\rho (T,H)/L^{2},
    \label{eq:two}
\end{equation} 
where $\rho (T,H)$ is the electrical resistivity.  Then
$\Theta_{1}^{\prime\prime}/h$ is a function of only $\omega/\omega_{peak}$,
and has a peak at $\omega = \omega_{peak}$.  When $\delta_{s}$ becomes
infinity, the field penetrates the system completely and
$\Theta_{1}^{\prime\prime}/h$ tends to zero.  In the opposite limiting case
($\delta_{s} \rightarrow 0$), the screening is complete and
$\Theta_{1}^{\prime\prime}/h = 0$.

First we consider the peak temperature $T_{peak}(H)$ of
$\Theta_{1}^{\prime\prime}/h$ at any $h$, where $T_{peak}(H) > T_{gl}(H)$. 
We assume that the response of the system is linear (Ohmic).  The current
density is of the order of $J_{peak} = ch/(4\pi L)$.  The resistivity $\rho
(T,H) = E(J_{peak})/J_{peak}$, which is independent of $J_{peak}$,
decreases with decreasing $T$ and tends to zero at $T_{gl}(H)$.  The peak
temperature $T_{peak}(H)$ can be determined from the matching condition
($\omega = \omega_{peak}$) given by Eq.(\ref{eq:two}): $T = T_{peak}^{(1)}$
for $f_{1} = 0.1$ Hz and at $T = T_{peak}^{(2)}$ for $f_{2} = 1$ kHz.  Then
we have a relation $\rho (T_{peak}^{(2)}, H)/\rho (T_{peak}^{(1)}, H) =
10^{4}$.  Since there is no shift of $T_{peak}(H)$ in
$\Theta_{1}^{\prime\prime}/h$ with $f$ ($0.1 \leq f < 1000$ Hz) in the
present work, it follows that the difference of peak temperatures defined
by $\Delta T = T_{peak}^{(2)} - T_{peak}^{(1)}$ should be less than 0.01 K.
Because of the sharp drop in resistivity with decreasing $T$, $T_{peak}(H)$
thus obtained is very close to but above $T_{gl}(H)$.  The value of
$T_{peak}(H)$ is independent of $h$ because Eq.(\ref{eq:two}) is
independent of $h$.  The resistivity is finite at $T = T_{peak}(H)$:
$\rho_{min} = \rho (T_{peak}, H) = 1.25\omega L^{2}/c^{2}$.  In summary,
$\Theta_{1}^{\prime\prime}/h$ vs $T$ has a peak at $T_{peak}(H)$ just above
$T_{gl}(H)$, which is independent of $h$.

Next we consider the peak temperature $T_{peak}(H)$ of
$\Theta_{1}^{\prime\prime}/h$ at any $h$, where $T_{peak}(H) < T_{gl}(H)$. 
We assume that the response of the system is highly nonlinear (non-Ohmic). 
The effective resistivity is defined by $\rho_{eff}(J_{peak}, T, H) =
E(J_{peak})/J_{peak}$, where $J_{peak} =ch/(4\pi L)$.  As a function of
$T$, $\rho_{eff}$ decreases with decreasing $T$ and tends to zero at a
temperature below $T_{gl}(H)$.  As a function of $J_{peak}$, $\rho_{eff}$
has a local minimum at $J_{peak} = (A\mu)^{1/\mu}$, and is equal to zero at
both $J_{peak} = 0$ and $J_{peak} = \infty$.  When $J_{peak}$ is smaller
than $(A\mu)^{1/\mu}$ [or $h < 4\pi L(A\mu)^{1/\mu}/c$],
$\rho_{eff}(J_{peak})$ decreases with decreasing $J_{peak}$ (or with
decreasing $h$).  The peak temperature $T_{peak}(H)$ can be estimated from
Eq.(\ref{eq:two}) with $\rho(T,H) = \rho_{eff}(J_{peak})$:
$\rho_{eff}(J_{peak}) = 1.25\omega L^{2}/c^{2}$ at $T = T_{peak}(H)$.  The
value of $J_{peak}$ (or $h$) is uniquely determined because of the relation
between $J_{peak}$ and $h$.  The peak temperature $T_{peak}(H)$, which is
lower than $T_{gl}(H)$, increases with decreasing $h$ and approaches the
line $T_{gl}(H)$ from the low-$T$ side.  In summary,
$\Theta_{1}^{\prime\prime}/h$ vs $T$ has a peak at $T_{peak}(H)$ just below
$T_{gl}(H)$ for sufficiently large $h$ satisfying the above condition.

The lines $H_{gl}$ and $H_{gl^{\prime}}$ intersect at a critical point
located at $T_{0}^{\prime} = 2.5$ K and $H_{0}^{\prime} = 160$ Oe.  The
line $H_{gl}$ line for $H > H_{0}^{\prime}$ no longer separates the
irreversible region of the vortex glass phase from the reversible region of
the vortex liquid phase.  In this sense, the vortex glass phase may connect
smoothly to the vortex liquid phase at the critical point at
$T_{0}^{\prime}$ and $H_{0}^{\prime}$ \cite{Menon2002}.  As shown in
Fig.~\ref{fig:three}, in fact the $T$ dependence of $\delta$ at $H \geq
200$ Oe is rather different from that at 150 Oe.  The difference $\delta$
at $H \geq 250$ Oe is almost equal to zero at any $T$, indicating the
reversibility of magnetization on the line $H_{gl}(T)$.

The line $H_{gl}$($T$) may correspond to the AT line of spin glasses
\cite{Almeida1978}.  Then it is expected that the relaxation time of Sn-MG,
$\tau$, diverges as a power law form $\tau = \tau_{0}[T/T_{gl}(H) -1]^{-x}$
\cite{Blatter1994}, as $T$ approaches the irreversibility line $T_{gl}(H)$
from the high-$T$ side.  Here $\tau_{0}$ is a characteristic relaxation
time and $x$ is a dynamic critical exponent.  Experimentally, no relaxation
time effect is observed in the $T$ and $f$ dependence of
$\Theta_{1}^{\prime}/h$ and $\Theta_{1}^{\prime\prime}/h$ on the line
$H_{gl}$($T$) (see Sec.~\ref{resultE}).  This result indicates that the
condition $\omega \tau \ll 1$ is satisfied, in spite of crossing the line
$H_{gl}(T)$.  Thus the power law form is not the case of the present
system.  What is the origin of $\tau$ satisfying the condition $\omega \tau
\ll 1$?  The absorption of the AC field is mainly governed by the hopping
of the vortices over the barriers between different metastable states.  The
depinning time $\tau$ for such intervalley transition is obtained from a
characteristic relaxation time $\tau_{0}$ by an Arrhenius form $\tau
\approx \tau_{0} \exp (U_{0}/k_{B}T)$, where $U_{0}$ is the activation
barrier energy between the different metastable states
\cite{Geshkenbein1991,Steel1992}.  Then $\tau$ smoothly changes as $T$
changes and $\omega \tau \ll 1$ on the line $H_{gl}(T)$.

In summary, it is concluded from the experimental results described above
that the properties of the vortex glass phase in Sn-MG are characterized as
follows, although most of them are not sufficiently understood at present. 
(i) The magnetization $M_{ZFC}$ deviates from $M_{FC}$ below $T_{gl}(H)$
for $H > H^{*}$.  (ii) The magnetization $M_{IR}$ deviates from $M_{TR}$
below $T_{gl}(H)$.  The $T$ dependence of $M_{IR}$ and $M_{TR}$ is very
similar to that of $M_{ZFC}$ and $M_{FC}$ in typical spin glass systems,
respectively \cite{Mydosh1993}.  (iii) The difference ($\delta - \Delta$)
takes a positive large value for the vortex glass phase, while it is equal
to zero for the vortex liquid phase and it takes a negative value in the
vortex lattice phase.  (iv) The relaxation time $\tau$ shows no behavior of
divergence on the line $T_{gl}(H)$.  (v) The non-Ohmic behavior is observed
in the vortex glass phase, while the Ohmic behavior is observed in the
vortex liquid phase.

\subsection{\label{disC}Antiferromagnetic short-range order}
There are several results supporting the existence of the AF short-range
order.  The first evidence is the $T$ dependence of the peak field $H_{p}$
as shown in the inset of Fig.  2.  The field $H_{p}$ is much higher than
$H_{c2}$ at the same $T$: $H_{p} \approx 3.2$ kOe and $H_{c2} \approx 490$
Oe at 3.0 K. The field $H_{p}$ increases with decreasing $T$ and may be
described by a power law form given by $H_{p} =
H_{p}^{*}(1-T/T_{p}^{*})^{\beta}$, where $\beta$ is an exponent, and
$H_{p}^{*}$ and $T_{p}^{*}$ are characteristic field and temperature,
respectively.  The least squares fit of the data of $H_{p}$ vs $T$ ($1 \leq
H \leq 3$ kOe) to the power law form yields the parameters $\beta = 0.41
\pm 0.02$, $T_{p}^{*} = 3.82 \pm 0.02$ K, and $H_{p}^{*} = 5.95 \pm 0.05$
kOe.  The exponent $\beta$ is close to $\beta = 1/2$ for the molecular
field theory.  The temperature $T_{p}^{*}$ is a little higher than $T_{c}$. 
The second evidence is the magnitude of the DC magnetic susceptibility:
$\chi = 3.45 \times 10^{-6}$ emu/g at $H = 1$ kOe and $T = 5$ K. This value
of $\chi$ is positive and is very different from that of pristine graphite
which is diamagnetic (see Sec.~\ref{resultC}) \cite{Heremans1994}.

What is the origin of the AF short-range order?  The existence of the AF
short-range order has been observed in other MG's such as Bi-MG
\cite{Suzuki2002b}.  In these MG's the metal layers would generate internal
stress inside the graphite lattice, leading to the break up of adjacent
graphene sheets into nanographites.  Fujita et al.  \cite{Fujita1996} and
Wakabayashi et al.  \cite{Wakabayashi1999} have theoretically suggested
that the electronic structures of finite-size graphene sheets depend
crucially on the shape of their edges (zigzag type and armchair type
edges).  Finite graphite systems having zigzag edges exhibit a special edge
state.  The corresponding energy bands are almost flat at the Fermi energy,
thereby giving a sharp peak in the density of states (DOS) at the Fermi
energy.  As a result, conduction electrons are localized around the zig-zag
edges of nanographites, forming localized magnetic moments.  Recently
Harigaya \cite{Harigaya2001} has theoretically predicted that the magnetism
in nanographites with zigzag edge sites depends on the stacking sequence of
nanographites.  The antiferromagntic spin alignment becomes energetically
favorable when the adjacent graphene sheets have a A-B type stacking
sequence along the $c$ axis, like pristine graphite \cite{Enoki2003}.  The
AF short-range order in nanographites may coexist with the
superconductivity occurring in Sn sheets even at low $T$ and low $H$.

\section{\label{conc}CONCLUSION} 
Sn-MG is a quasi-2D superconductor, showing a superconducting transition at
$T_{c} = 3.75$ K for $H = 0$.  The $H$-$T$ diagram of Sn-MG consists of
four lines ($H_{ag}$, $H_{gl}$, $H_{gl^{\prime}}$, and $H_{al}$) separating
vortex lattice phase, vortex glass phase, and vortex liquid phase.  These
lines merge into the multicritical point located at $T^{*} = 3.4$ K and
$H^{*} = 40$ Oe.  The irreversibility lines $H_{gl}$ and $H_{gl^{\prime}}$
intersect at a critical point at $T_{0}^{\prime}$ = 2.5 K and
$H_{0}^{\prime} = 160$ Oe.  The line $H_{gl}$ has an AT-power law form,
while the line $H_{gl^{\prime}}$ has a GT-power law form.  The slow
dynamics and nonlinearity in the vortex glass phase is an analogy to that
in the spin glass phase observed in 3D Ising spin glasses.  Further studies
will be required to understanding the nature of the vortex glass phase from
measurements of resistivity (the nonlinear relation of current-voltage) and
the absorption and dispersion at frequencies ($f \gg 1$ kHz).

\ack
The work at Binghamton (M.S. and I.S.S.) was supported by the Research
Foundation of SUNY-Binghamton (contract number 240-9522A).  The work at
Osaka (J.W.) was supported by the Ministry of Education, Science, Sports
and Culture, Japan [the grant for young scientists (No.  70314375)] and by
Kansai Invention Center, Kyoto, Japan.

\end{document}